\documentclass[twocolumn,showpacs,aps,prb]{revtex4}% Physical Review Letters
\usepackage{graphicx}% Include figure files
\usepackage{dcolumn}% Align table columns on decimal point
\usepackage{bm}% bold math

\begin{document}

\title{Fine-structure splitting of exciton states in quantum dot molecules: symmetry and tunnel-coupling effects}

\author{Dong Xu}
%\email[Electronic address: ]{d-xu03@mails.tsinghua.edu.cn}
\affiliation{%
Department of Physics, Tsinghua University, Beijing 100084, People's Republic
of China}
\author{Nan Zhao}
%\email[Electronic address: ]{d-xu03@mails.tsinghua.edu.cn}
\affiliation{%
Department of Physics, Tsinghua University, Beijing 100084, People's Republic
of China}
\author{Jia-Lin Zhu}
\email[Electronic address: ]{zjl-dmp@tsinghua.edu.cn}
\affiliation{%
Department of Physics, Tsinghua University, Beijing 100084, People's Republic
of China}
\date{\today}% It is always \today, today,
             %  but any date may be explicitly specified

\begin{abstract}
Exciton levels and fine-structure splitting in laterally-coupled quantum dot
molecules are studied. The electron and hole tunneling energies as well as the
direct Coulomb interaction are essential for the exciton levels. It is found
that fine-structure splittings of the two-lowest exciton levels are contributed
from the intra- and interdot exchange interactions which are greatly influenced
by the symmetry and tunnel-coupling between the two dots. As the inter-dot
separation is reduced, fine-structure splitting of the exciton ground state is
largely increased while those of the excited states are decreased. Moreover,
the dependence of the fine-structure splitting in quantum dot molecules on the
Coulomb correlation is clearly clarified.
\end{abstract}
% PACS, the Physics and Astronomy.
                             % Classification Scheme.
%\keywords{Suggested keywords}%Use showkeys class option if keyword
                              %display desired
\maketitle

\section{Introduction}
During the last few years, there are intensive studies on the optical
properties of single semiconductor quantum dots (QDs), since they exhibit
atom-like energy spectrum and sharp lines in photoluminescence as well as the
advantage that they can be conveniently manipulated by the external fields.
Semiconductor QDs have been been demonstrated as one of the candidates for the
single-photon or entangled two-photon sources, which make them very attractive
for applications in the fields of quantum cryptography and quantum
teleportation.~\cite{michler1,michler2,Akopian,Pelton,Santori1,Santori2,Stevenson}
In the first proposal for a QD-based source of polarization entangled photon
pairs, a necessary condition is that the intermediate monoexciton states for
the biexciton radiative decay are energetically degenerate.~\cite{Benson}
However, the III-V self-assembled semiconductor QDs tend to be elongated along
the $[\bar{1}10]$ crystal axis and the monoexciton states are split by the
anisotropic electron-hole exchange
interaction.~\cite{Gammon,Bayer,Ivchenko,Takagahara,Bester1,Seguin}
Consequently, much effort is devoted to the reduction of the fine-structure
splitting of the intermediate exciton states, e.g., thermal annealing and
external field
tuning.~\cite{Young,Tartakovskii,Langbein,Greilich,Stevenson2,Kowalik,Kowalik1}

%Advances in single-dot spectroscopy revealed that photoluminescence lines of
%III-V QDs show doublet structures as well as linear polarization.

Neglecting the electron-hole exchange interaction, the exciton ground state is
four-fold degenerate. For In$_{x}$Ga$_{1-x}$As QDs with
$\mathrm{C}_{\mathrm{2v}}$ or lower symmetry, however, the electron-hole
exchange interaction splits the exciton ground state into bright and dark
doublets, which are separated by about few hundred $\mu$eV.~\cite{Bayer3}
Furthermore, the bright doublets are split into two linearly polarized states
separated by about tens of $\mu$eV.~\cite{Bayer2} At zero external field, the
fine-structure splitting of the bright doublet mainly depends on the anisotropy
and size of QDs.~\cite{Bester1,Takagahara,Ivchenko} If a magnetic field is
applied in a Voigt configuration, the bright states are mixed with the dark
states through the Zeeman term and the dark exciton states become optically
active.~\cite{Bayer3,Bayer2} Due to the Zeeman splitting induced by the
magnetic field, the fine-structure splitting of bright doublet could be tuned
to zero by the magnetic field.~\cite{Stevenson,Stevenson2} In addition, thermal
annealing~\cite{Langbein,Tartakovskii} and an in-plane electric
field~\cite{Kowalik,Kowalik1} could lead to a significant reduction of the
fine-structure splitting.

Recently, with the development of high-quality QD structures, it is possible to
fabricate either vertically- or laterally-coupled self-assembled QDs, namely
``quantum dot molecule'' (QDM).~\cite{Bayer4,Yugova,Gerardot,Unold,Beirne} The
exciton ground states exhibit fine structures induced by the electron and hole
tunnel-coupling between the two dots.~\cite{Ortner7} The exciton states and
interdot coupling could be manipulated by the electric field in either
vertically- or laterally-coupled
QDM.~\cite{Ortner6,Krenner7,Beirne,szafran,szafran1,Bester19,zhu} Significant
Stark effect~\cite{Ortner6,Beirne} and pronounced anticrossing of different
excitonic transitions~\cite{Krenner7} have been observed in the
photoluminescence (PL) spectra. According to several numerical
calculations,~\cite{Bester9,zhu,szafran} interdot coupling as well as the
symmetry between the two dots strongly influence the exciton levels and optical
properties.

Previously, we found that fine-structure splitting of the exciton ground state
in laterally-coupled QDM could be tuned to zero by applying an in-plane
electric field of only few kV/cm, which might overcome the deficiency in a
single QD.~\cite{Zhu2} Polarization entangled photon pairs might be prepared in
a QDM through the biexciton radiative decay. Thus clarifying the mechanism of
the variation of fine-structure splitting in QDMs with various interdot
separations is very important. In this paper, we study the symmetry and
tunnel-coupling effects on the fine-structure splitting of exciton states in
QDMs. Exciton levels of symmetric and nonsymmetric QDMs without the
electron-hole exchange interaction are given, and we note that the exciton
level spacing is much larger than the electron-hole exchange energies.
Fine-structure splittings of the two-lowest exciton levels are contributed from
the intra- and interdot exchange interactions. Symmetry and tunnel-coupling in
QDM strongly influence the exciton envelope functions, and therefore it is
interesting that intra- and interdot exchange interactions might be strongly
dependent on the symmetry and tunnel-coupling. Moreover, the variations of the
fine-structure splittings of exciton ground and excited states are compared and
discussed.

%The primary purposes of this paper are to study the exciton levels and
%fine-structure splitting in laterally-coupled QDM, and to provide a new way of
%controlling the exciton fine-structure splitting. In typical tunnel-coupled
%QDs, the electron and hole tunneling energies and the direct Coulomb
%interaction are essential for the exciton envelop functions. In order to
%compare with the experimental results, we choose size of QDs resembling those
%grown in the experiments and calculated the exciton fine-structure splitting in
%both single QD and QDM. Moreover, the dependence of the fine-structure
%splitting in QDM on the Coulomb correlation is shown. In addition, considering
%the size fluctuations in really grown QD samples, we investigate the
%fine-structure splitting in coupled nonidentical QDs.

In Sec.~II, a microscopic theory of exciton levels and fine-structure splitting
in QDM is formulated. In Sec.~III, exciton levels in QDM without the exchange
interaction are given and discussed. Fine-structure splittings of few low-lying
bright exciton states are shown for both symmetric and nonsymmetric QDMs. The
intra- and interdot parts of the splittings of the two-lowest exciton levels
are given and compared. The effects of the symmetry and tunnel-coupling as well
as the Coulomb correlation are clearly clarified. Finally, the results are
summarized in Sec. IV.

%We formulate a theory of exciton fine-structure splitting in laterally QDM
%taking into account the Coulomb correlation effect.

\section{Description of Exciton Levels and Fine-Structure Splitting}
The exciton fine-structure splitting in the semiconductor QDs is contributed
from the electron-hole exchange interaction. The study of the electron-hole
exchange interaction requires a clear representation of the spin states of the
exciton. In a bulk III-V direct-gap semiconductor, the valence-band edge has
$\Gamma_{8}$ symmetry ($J=3/2$) and the conduction-band edge has $\Gamma_{6}$
symmetry. For the flat InGaAs quantum dots investigated in this paper, the
light- ($J_z=\pm1/2$) and heavy-hole ($J_z=\pm3/2$) bands are split by several
tens of meV due to the strain introduced by the lattice mismatch. Actually,
according to an empirical tight-binding calculation,~\cite{Sheng} the
proportion of heavy-hole component in the hole ground state of a flat InGaAs QD
is as large as 98.2\%. Therefore, it is reasonable that the light-hole and
spin-orbit-split $J=1/2$ valence band could be neglected since we mainly focus
on the fine-structure splittings of few low-lying exciton states. The exciton
state is composted of four combinations of the valence band and the conduction
band, i.e.,
\begin{eqnarray}\label{envelope function}
|X\rangle=\sum_{m,s}\sum_{r_e,r_h}
\psi_{ms}(r_e,r_h)a^{\dag}_{c_sr_e}a_{v_mr_h}|0\rangle,
\end{eqnarray}
where the Wannier function representation of the creation and annihilation
operators is used, $m$ and $s$ are the $z$ component of the angular momentum of
the heavy-hole valence band and the conduction band, respectively, and
$\psi_{ms}(r_e,r_h)$ is the envelope function. The $z$ component of exciton
spin for state $(m,s)$ is $s-m$. From the selection rule, the spin $\pm$1
exciton states might be optically active depending on the orbital envelop
functions while the spin $\pm$2 states are optically inactive, irrespective of
the orbital envelop functions. The eigenvalue equation for $\psi_{ms}$ is given
as {\setlength\arraycolsep{2pt}
\begin{eqnarray}\label{eigenvalue equation}
\sum_{m's'r'_er'_h}[H_1&+&V_{\mathrm{ex}}(c_sr_e,v_{m'}r'_h;c_{s'}r'_e,v_mr_h)]\psi_{m's'}(r'_e,r'_h)
\nonumber\\&&=E\psi_{ms}(r_e,r_h),
\end{eqnarray}}
with the spin-independent part
\begin{eqnarray}\label{spin-independent Hamiltonian}
H_1=\delta_{r_er'_e}\delta_{r_hr'_h}\delta_{s's}\delta_{m'm}[\frac{p^2_e}{2m_e}+U_e(r_e)
\nonumber\\+\frac{p^2_h}{2m_h}+U_h(r_h)-\frac{e^2}{\epsilon|r_e-r_h|}],
\end{eqnarray}
where $U_e$ ($U_h$) is the confinement potential for the conduction (valence)
band electron. The electron-hole exchange interaction $V_{\mathrm{ex}}$ can be
approximated as {\setlength\arraycolsep{2pt}
\begin{eqnarray}\label{exchange interaction}
&&V_{\mathrm{ex}}(c_sr_e,v_{m'}r'_h;c_{s'}r'_e,v_mr_h)\nonumber\\
&&\approx\delta_{r_er_h}\delta_{r'_er'_h}[\delta_{r_er'_e}V(c_sr_e,v_{m'}r_e;c_{s'}r_e,v_mr_e)
\nonumber\\&&+(1-\delta_{r_er'_e})V(c_sr_e,v_{m'}r'_e;c_{s'}r'_e,v_mr_e)],
\end{eqnarray}}
The first (second) term of Eq.(\ref{exchange interaction}) is the so-called
short-range (long-range) exchange interaction. The long-range term can be
further approximated through the multipole expansion
\begin{eqnarray}\label{dipole dipole}
V(c_sr_e,v_{m'}r'_e;c_{s'}r'_e,v_{m}r_e)\nonumber\\\approx\vec{\mu}_{c_s,v_m}\frac{[1-3n\cdot^tn]}{|r_e-r'_e|^3}\vec{\mu}_{v_{m'},c_{s'}}
\end{eqnarray}
with
\begin{eqnarray}\label{dipole}
&&n=\frac{r_e-r'_e}{|r_e-r'_e|}\nonumber\\
&&\vec{\mu}_{c_s,v_m}=e\int d^3r \phi^{\ast}_{c_sR}(r)\vec{r}\phi_{v_mR}(r)
\end{eqnarray}
where $\phi_{cs(vm)R}(r)$ is a Wannier function localized at the site $R$. The
matrix element of $V_{\mathrm{ex}}$ is given as~\cite{Takagahara}
\begin{eqnarray}\label{four component}
V_{\mathrm{anal}}(ms,m's')\int d^3r \psi^{\ast}_{ms}(r,r)\psi_{m's'}(r,r)\nonumber\\
+\int d^3r~\mathrm{div}_r(\psi^{\ast}_{ms}(r,r)\vec{\mu}_{c_s,v_m})\nonumber\\
\times \mathrm{div}_r[\int
d^3r'\psi_{m's'}(r',r')\frac{\vec{\mu}_{v_{m'},c_{s'}}}{|r-r'|}]
\end{eqnarray}
The analytical part $V_{\mathrm{anal}}$ could be written in the matrix form as
\begin{eqnarray}\label{analytical part}
&&V_{\mathrm{anal}}(ms,m's')\nonumber\\
&&=(E^S_X-8\pi/3\mu^2)\left( \begin{array}{cccc}
1&0&0&0\\
0&0&0&0\\
0&0&0&0\\
0&0&0&1
\end{array}\right)
\end{eqnarray}
with $(m,s)$ and $(m',s')$ in the order, $(\frac{3}{2},\frac{1}{2})$,
$(\frac{3}{2},-\frac{1}{2})$, $(-\frac{3}{2},\frac{1}{2})$, and
$(-\frac{3}{2},-\frac{1}{2})$. The definition of the parameters $E^S_X$ and
$\mu$ can be found in Ref.~\onlinecite{Takagahara}. The second term of
Eq.~\ref{four component}, i.e., the nonanalytic term, has a form given by
\begin{eqnarray}\label{nonanalytical part}
\left( \begin{array}{cccc}
*&0&0&*\\
0&0&0&0\\
0&0&0&0\\
{*}&0&0&*
\end{array}\right)
\end{eqnarray}
where $\ast$ indicates nonzero matrix elements. In anisotropic QDs, the
nondiagonal elements of Eq.~\ref{nonanalytical part} are nonzero, and thus the
exciton spin $\pm1$ states are split into two linearly polarized states. The
fine-structure splitting of the doublet is determined by the nondiagonal
elements of Eq.~\ref{nonanalytical part}.

Similar to the assumption in Ref.~\onlinecite{Ivchenko}, we use an in-plane
anisotropic potential to model a single QD, and the two dots are aligned along
the $x$ axis
\begin{eqnarray}\label{potential}
&&U_{e(h)}=\nu_{e(h)}[\theta(\frac{b_0}{2}-|y_{e(h)}|)\theta(\frac{a_0}{2}-|x_{e(h)}+\frac{d+a_0}{2}|)
\nonumber\\&&{}+\theta(\frac{b_1}{2}-|y_{e(h)}|)\theta(\frac{a_1}{2}-|x_{e(h)}-\frac{d+a_1}{2}|)]
\end{eqnarray}
where the two dots with lateral size $a_i$ and $b_i$ for the $i$th dot are
separated by distance $d$, and $\nu_{e}$ ($\nu_{h}$) is the conduction
(heavy-hole valence) band offset. The exciton envelope function in QDM can be
expanded using the Hermite polynomials as
\begin{eqnarray}\label{hermite expansion}
&&\psi(r_e,r_h)=\sum_{i,j,m,n}C_{ijmn}A_{ijmn}\nonumber\\ &&\times u_i(\alpha
x_e)u_j(\alpha y_e)\mathrm{exp}[-\frac{1}{2}\alpha^2(x^2_e+y^2_e)]\nonumber\\
&&\times u_m(\alpha x_h)u_n(\alpha
y_h)\mathrm{exp}[-\frac{1}{2}\alpha^2(x^2_h+y^2_h)]
\end{eqnarray}
where $u_i(x)$ is the Hermite polynomial, $A_{ijmn}$ is the normalization
coefficient for the Hermite polynomials, $C_{ijmn}$ is the expansion
coefficient, and $\alpha$ is a variational parameter. Since we study the flat
QDs in this paper, the two-dimensional approximation is assumed in the
calculation. About 4800 lowest-energy envelope basis are taken into account in
the diagonalization of the spin-independent matrix $H_1$ to ensure the
convergence of the calculation. We note that the matrix elements of the Coulomb
interaction could be analytically obtained.~\cite{Ugajin} The fine-structure
splitting then could be obtained through the perturbation calculation of the
exchange interaction $V_{\mathrm{ex}}$ since the exchange energy (tens of
$\mu$eV) is much less than the exciton level spacing (tens of meV).

\section{Numerical Calculation and Discussion}

\begin{figure}
\begin{center}
\includegraphics*[angle=0,width=0.43\textwidth]{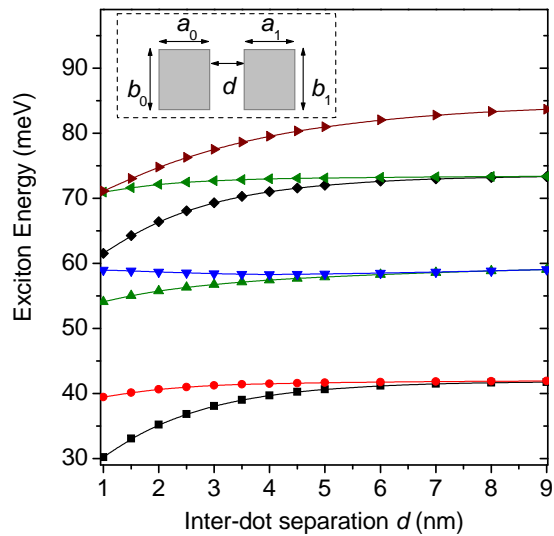}
\caption{(Color online) Low-lying exciton levels without the exchange
interaction as functions of the interdot separation $d$ for two coupled
identical dots with $a_0=a_1=16$ nm and $b_0=b_1=20$ nm. Inset: schematic
illustration of a laterally-coupled QDM.} \label{FIG:level identical dots}
\end{center}
\end{figure}

\begin{figure}
\begin{center}
\includegraphics*[angle=0,width=0.43\textwidth]{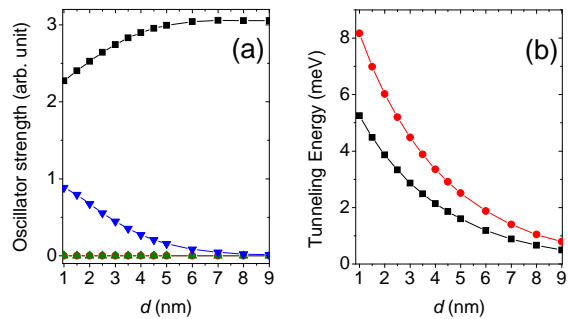}
\caption{(Color online) (a) Oscillator strengths of the four-lowest exciton
states of QDM in Fig. 1 and (b) tunneling energies for the electron (circle)
and hole (square), respectively.} \label{FIG:os tunnel}
\end{center}
\end{figure}

%\begin{figure}
%\begin{center}
%\includegraphics*[angle=0,width=0.43\textwidth]{Fig3.eps}
%\caption{Schematic diagram of exciton fine structures induced by the exchange
%interaction in coupled identical QDs. The optically active and inactive states
%are shown by solid and dash lines, respectively.} \label{FIG:schematic ps}
%\end{center}
%\end{figure}

In QDM, the exciton levels are strongly affected by the tunneling of the
electron and hole as well as the direct Coulomb interaction. The tunneling
energies of the electron and hole in the strong coupling region are much larger
than the electron-hole exchange energies. The values of the material parameters
used in the calculation are $\nu_h=81$ meV, $\nu_e=124$ meV, $m_e$=0.034 $m_0$,
$m_h$=0.053 $m_0$, $\mu=6$ $e\mathrm{\AA}$, $\epsilon_r$=14.

\subsection{Symmetric Quantum Dot Molecule}

In Fig.~\ref{FIG:level identical dots}, low-lying exciton levels without the
electron-hole exchange interaction for two coupled identical dots with
$a_0=a_1=16$ nm and $b_0=b_1=20$ nm are shown as functions of the interdot
separation $d$. At lager $d$ ($>8$ nm), the coupling between the two dots is
very weak and the exciton ground state energy is much close to that of a single
isolated dot with $a=16$ nm and $b=20$ nm. As the separation $d$ decreases, the
interdot coupling is enhanced and the energy levels are split. We note that the
four lowest exciton states are mainly composed of the electron and hole ground
state in individual dots.~\cite{Bester9} Their oscillator strengths are shown
as functions of $d$ in Fig.~\ref{FIG:os tunnel}(a). As $d$ decreases,
oscillator strength of the ground state decreases while that of the fourth
state increases. The second and third states are always optically inactive, and
the fourth state becomes optically inactive as $d>7.0$ nm. Fig.~\ref{FIG:os
tunnel}(b) shows the tunneling energies of the electron and hole, which are
calculated by the numerical diagonalization of the single particle Hamiltonian.
As $d$ decreases from 9.0 nm to 1.0 nm, the absolute values of the tunneling
energies for the electron (hole) increase from 0.79 meV (0.50 meV) to 8.17 meV
(5.25 meV).

Exciton levels of QDM without the exchange interaction are each four-fold
degenerate and the level spacing is about several meV. If the exchange
interaction is included, each of exciton levels is split into fine-structures.
One doublet is composed of exciton spin $\pm2$ states, and the other is
composed of exciton spin $\pm1$ states. For symmetric QDMs in the strong
coupling region, the spin $\pm1$ doublets of the exciton ground and fourth
states as shown in Fig.~\ref{FIG:level identical dots} are optically active and
could be observed in the PL and PL-Excitation spectra.

In Fig.~\ref{FIG:ps identical dots}, fine-structure splittings of the exciton
ground ($\delta_1$) and fourth ($\delta_4$) states in symmetric QDMs with
$a_1=a_0=16$ nm and $b_1=b_0=20$ nm are shown as functions of $d$. At larger
$d$, the fine-structure splitting $\delta_1$ is negative, e.g., $\delta_1=-35$
$\mu$eV at $d=9.0$ nm. As $d$ decreases, $\delta_1$ increases from negative
values to positive values. At $d\approx4.4$ nm, $\delta_1$ is zero. However,
$\delta_4$ decreases from zero at larger $d$ to $-150 \mu$eV at $d=1.5$ nm.
Although tunnel-coupling induced splitting is negligible at $d=9.0$ nm as shown
in Fig.~\ref{FIG:level identical dots}, fine-structure splitting of the exciton
ground state is much different from that of the exciton ground state ($-66$
$\mu$eV) in a single isolated dot with the same size.

According to Eq.~(\ref{four component}), fine-structure splitting of the
exciton ground state can be approximated in the first-order as
\begin{eqnarray}\label{ps perturbe}
\delta_1 = 2\sum_{r_e\neq
r'_e}\psi_{1,s-m=1}(r_e,r_e)\psi_{1,-1}(r'_e,r'_e)V_{\mathrm{ex}}\nonumber\\
=\lambda_{1,\mathrm{intra}}+\lambda_{1,\mathrm{inter}}
\end{eqnarray}
with
\begin{eqnarray}\label{inter intra}
\lambda_{1,\mathrm{inter}}=2\sum_{r_e\neq r'_e,x_ex'_e<0}\psi_{1,1}(r_e,r_e)\psi_{1,-1}(r'_e,r'_e)V_{\mathrm{ex}},\\
\lambda_{1,\mathrm{intra}}=2\sum_{r_e\neq
r'_e,x_ex'_e>0}\psi_{1,1}(r_e,r_e)\psi_{1,-1}(r'_e,r'_e)V_{\mathrm{ex}}.
\end{eqnarray}
where $\psi_{i,s-m}(r_e,r_h)$ is the $i$th exciton eigenfunction of $H_1$ with
exciton spin $z$ component $s-m$, $\lambda_{1,\mathrm{intra}}$ is indeed the
long-range exchange interaction within individual dots, and
$\lambda_{1,\mathrm{inter}}$ is that between the two dots (namely the
F\"{o}rster interaction~\cite{Nazir}). Fig.~\ref{FIG:ps identical dots}(b)
shows $\lambda_{1,\mathrm{inter}}$ and $\lambda_{1,\mathrm{intra}}$ as
functions of $d$. $\lambda_{1,\mathrm{inter}}$ remains about 20 $\mu$eV while
$\lambda_{1,\mathrm{intra}}$ is greatly changed from $-54$ $\mu$eV at $d=9.0$
nm to $50$ $\mu$eV at $d=1.0$ nm.

\begin{figure*}
\begin{center}
\includegraphics*[angle=270,width=0.9\textwidth]{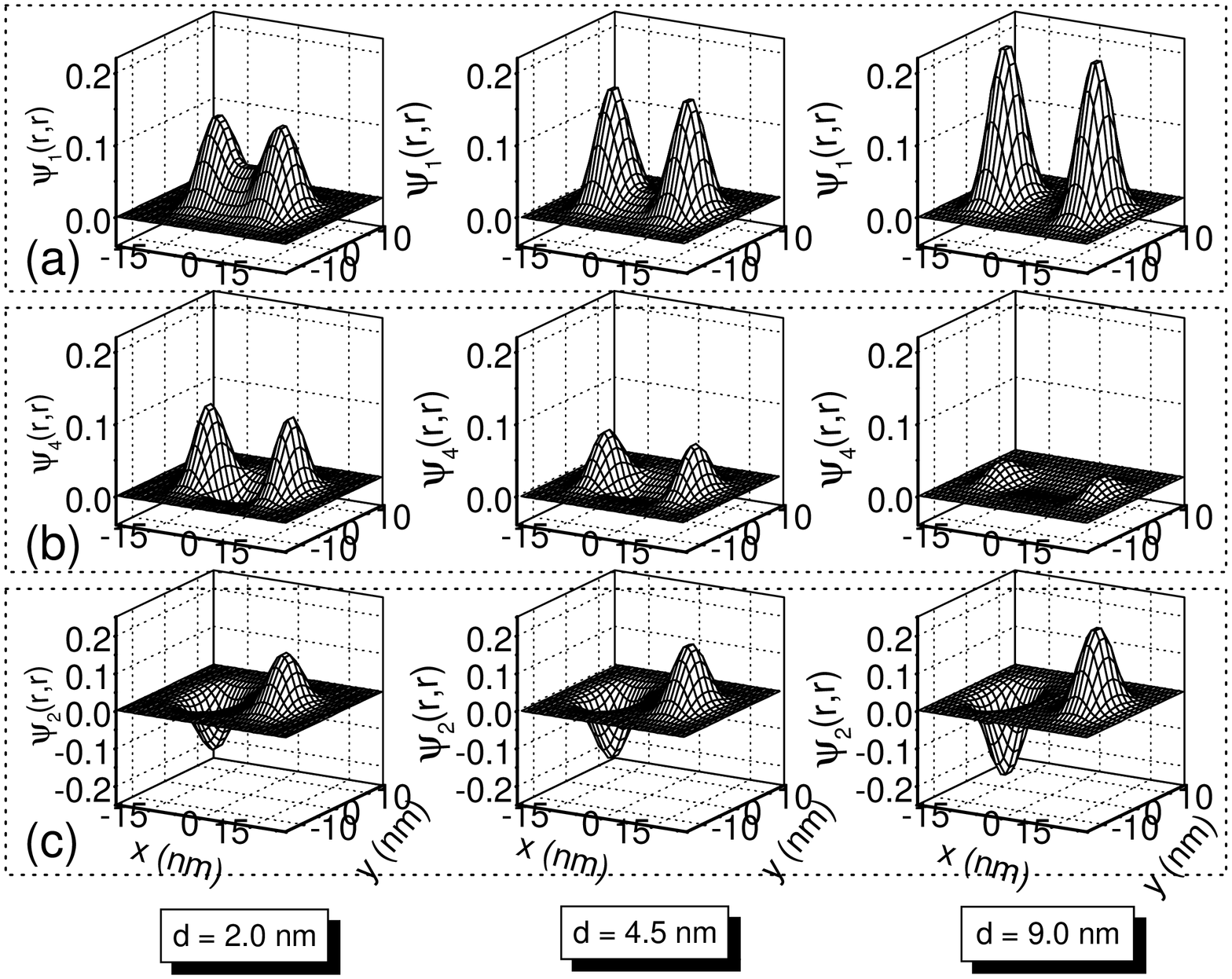}
\caption{(a) $\psi_1(r_e,r_h=r_e)$, (b) $\psi_4(r_e,r_h=r_e)$, and (c)
$\psi_2(r_e,r_h=r_e)$ of QDM in Fig. 1 for $d=2.0$, $4.5$, and $9.0$ nm,
respectively.} \label{FIG:pola dis 2d}
\end{center}
\end{figure*}

As discussed in Sec.~II, $\psi(r_e,r_h=r_e)$ is directly related to the
calculation of the exchange interaction, and therefore it is important for the
study of the variation of fine-structure splitting. In Figs.~\ref{FIG:pola dis
2d}(a) and \ref{FIG:pola dis 2d}(b), $\psi_1(r_e,r_h=r_e)$ and
$\psi_4(r_e,r_h=r_e)$ are plotted, respectively. $\psi_1(r_e,r_e)$ is composed
of two symmetrically-coupled $s$ state-like functions at larger $d$. For two
dots without the tunnel-coupling,~\cite{Nazir} the F\"{o}rster interaction
(i.e. $\lambda_{1,\mathrm{inter}}/2$) is simply enhanced as $d$ is reduced.
However, if the tunnel-coupling can not be neglected as shown in
Fig.~\ref{FIG:pola dis 2d}(a), amplitudes of two $s$ state-like functions are
reduced at smaller $d$, which largely compensates the enhancement of
$\lambda_{1,\mathrm{inter}}$ at smaller $d$ in the case of neglecting the
tunnel-coupling. The consequence is that $\lambda_{1,\mathrm{inter}}$ remains
almost constant in the range of few nanometer interdot separations, as shown in
Fig.~\ref{FIG:ps identical dots}(b). Moreover, two $s$ state-like functions are
strongly overlapped at smaller $d$ and the anisotropic shape of
$\psi_1(r_e,r_h=r_e)$ in individual dots is largely changed. That is why
$\lambda_{1,\mathrm{intra}}$ is greatly varied in the strong coupling region.
$\psi_4(r_e,r_e)$ is more complicated and there are two nodes along the $x$
axis. It almost disappears at larger $d$ and is largely enhanced at smaller
$d$. The second and third states are always optically inactive.
$\psi_2(r_e,r_h=r_e)$, as shown in Fig.~\ref{FIG:pola dis 2d}(c) for an
example, is completely antisymmetric.

\begin{figure}
\begin{center}
\includegraphics*[angle=0,width=0.43\textwidth]{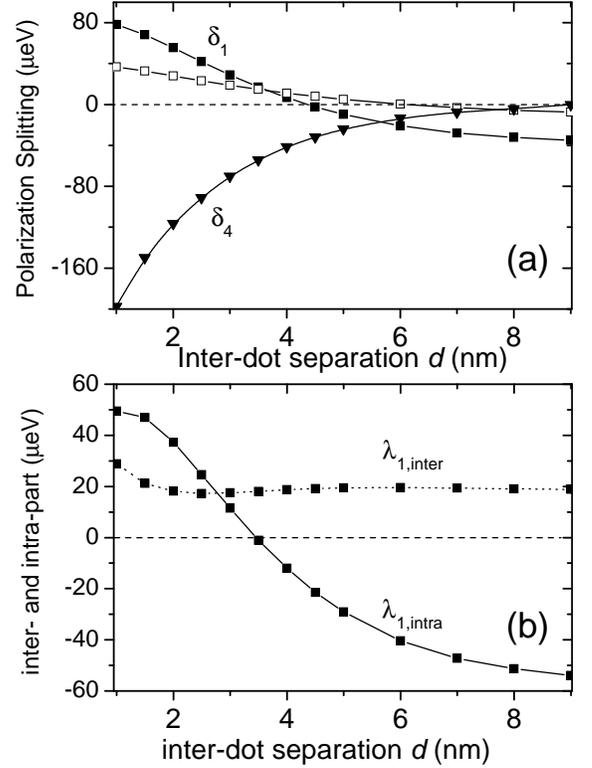}
\caption{(a) Fine-structure splitting of the exciton ground state with (square)
and without (open square) the Coulomb interaction, and of the fourth exciton
state (triangle) of QDM in Fig. 1; (b) $\lambda_{1,\mathrm{intra}}$ and
$\lambda_{1,\mathrm{inter}}$ as functions of $d$.} \label{FIG:ps identical
dots}
\end{center}
\end{figure}

\begin{figure}
\begin{center}
\includegraphics*[angle=0,width=0.5\textwidth]{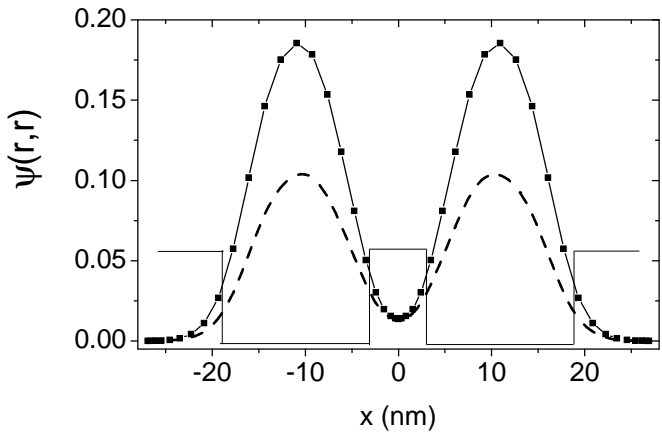}
\caption{$\psi(r_e,r_h=r_e)$ along the $y=0$ axis for the exciton ground state
in QDM of Fig. 1 at $d=6.0$ nm, with (square) and without (dash line) the
Coulomb interaction.} \label{FIG:polarizatioin dis 1d}
\end{center}
\end{figure}

The Coulomb correlation between the electron and hole is important for InGaAs
QDs with sizes that are comparable to or much larger than the exciton Bohr
radius. The exciton envelop function will be greatly changed by the Coulomb
correlation and thus the exciton fine-structure splitting could be largely
influenced. Fine-structure splitting of the exciton ground state in a single QD
($a_0=16$ nm,$b_0=20$ nm) without the Coulomb correlation is $-39$ $\mu$eV, the
deviation of which from that including the Coulomb correlation is as large as
$41\%$. In QDMs, the Coulomb correlation is also very important. Fine-structure
splitting of the exciton ground state without the Coulomb correlation in the
same QDM is compared with that taking into account the Coulomb correlation in
Fig.~\ref{FIG:ps identical dots}. It can be seen that the exciton
fine-structure splitting is much different between those with and without the
Coulomb correlation. For instance, at $d=6.0$ nm, $\delta_1\approx 0$ for the
case of independent electron and hole, while $\delta_1=-21 \mu$eV for the case
of correlated electron and hole. $\psi_1(r_e,r_h=r_e)$ with and without the
Coulomb correlation is shown along the $y=0$ axis in
Fig.~\ref{FIG:polarizatioin dis 1d} for $d=6.0$ nm, respectively.
$\psi_1(r_e,r_h=r_e)$ in the correlated case is obviously larger than that in
the independent one. That is why the exchange interaction and $\delta_1$ are
largely changed by the Coulomb correlation in QDMs.

%The numerically-calculated value of the fine-structure splitting $-66 \mu$eV in
%single QD with $a=16$ nm, $b=20$ nm is consistent with the experimental results
%of InGaAs QDs with similar sizes.~\cite{Bayer2} Larger value of the
%fine-structure splitting might arise from QDs with stronger anisotropy. For
%instance, QDM in Fig.~\ref{FIG:ps identical dots} would become a single QD
%with $a=32$ nm and $b=20$ nm as $d$ approaches zero and the polarization
%splitting would be larger than 80 $\mu$eV. Although it is now possible to
%fabricate laterally coupled InGaAs QDs,~\cite{Beirne} fine-structure splitting of
%the exciton states in QDM has not yet been reported in experiments.

\begin{figure}
\begin{center}
\includegraphics*[angle=0,width=0.43\textwidth]{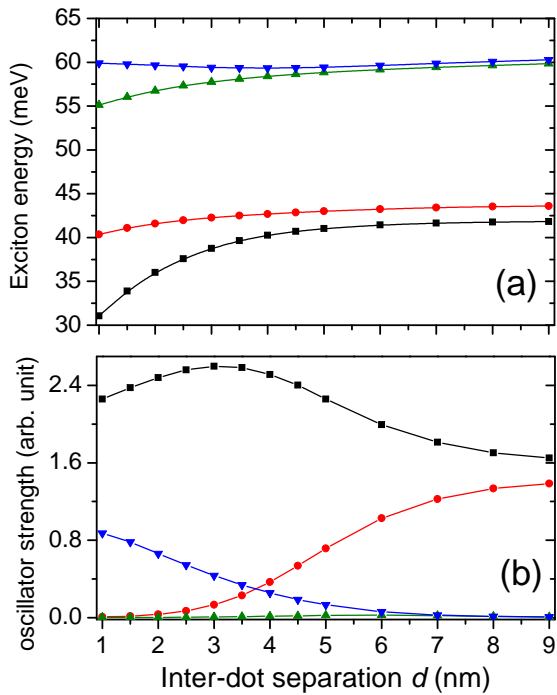}
\caption{(Color online) (a) The four-lowest exciton levels without the exchange
interaction and (b) the corresponding oscillator strengths as functions of $d$
for two coupled nonidentical dots with $a_0=a_1=16$ nm, $b_0=19$ nm, and
$b_1=20$ nm.} \label{FIG:level nonidentical dots}
\end{center}
\end{figure}

\subsection{Nonsymmetric Quantum Dot Molecule}
In experimental growth condition, there is size fluctuation in self-assembled
quantum dots and it may be difficult to obtain two coupled identical dots. Thus
it is important to study the fine-structure splitting in nonsymmetric QDMs. In
Fig.~\ref{FIG:level nonidentical dots}(a), the four lowest exciton levels
without the exchange interaction for two coupled nonidentical dots with
$a_0=a_1=16$ nm, $b_0=19$ nm, and $b_1=20$ nm are shown as functions of the
interdot separation $d$. As $d>8.0$ nm, the energies of the exciton ground and
second states are much close to the exciton ground state energies of the
isolated first and zeroth QDs, respectively. As the separation $d$ decreases,
the ground state energy becomes lower. Their oscillator strengths are shown as
functions of $d$ in Fig.~\ref{FIG:level nonidentical dots}(b). The oscillator
strength of the ground state first increases and then decreases while that of
the fourth state increases monotonically as $d$ decreases. The third state
remains almost optically inactive, and the fourth state becomes optically
inactive as $d>6.0$ nm. In contrast to the identical case as shown in
Fig.~\ref{FIG:os tunnel}(a), the second level is optically active at larger $d$
and becomes optically inactive as $d<2.0$ nm.

\begin{figure}
\begin{center}
\includegraphics*[angle=0,width=0.43\textwidth]{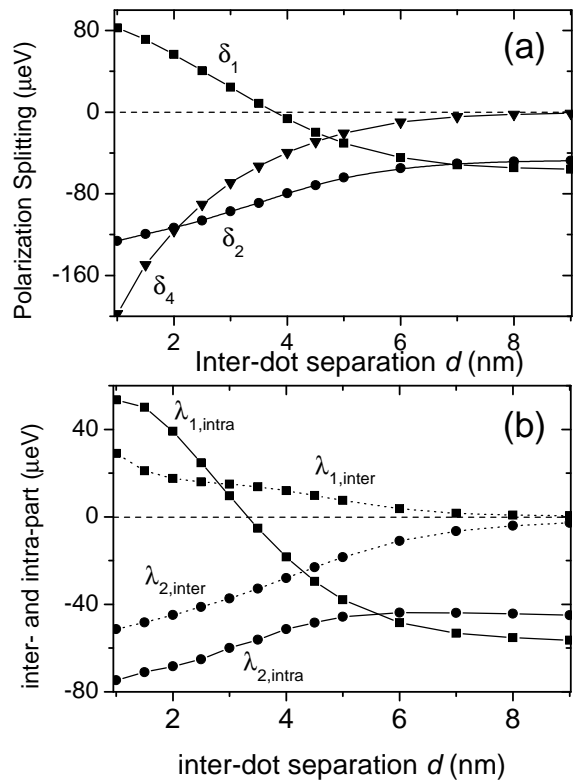}
\caption{(a) Fine-structure splitting of the ground (square), second (circle)
and fourth (triangle) states of QDM in Fig. 6; (b)
$\lambda_{1,\mathrm{intra}}$, $\lambda_{1,\mathrm{inter}}$,
$\lambda_{2,\mathrm{intra}}$, and $\lambda_{2,\mathrm{inter}}$ as functions of
$d$.} \label{FIG:ps nonidentical}
\end{center}
\end{figure}

\begin{figure*}
\begin{center}
\includegraphics*[angle=270,width=0.9\textwidth]{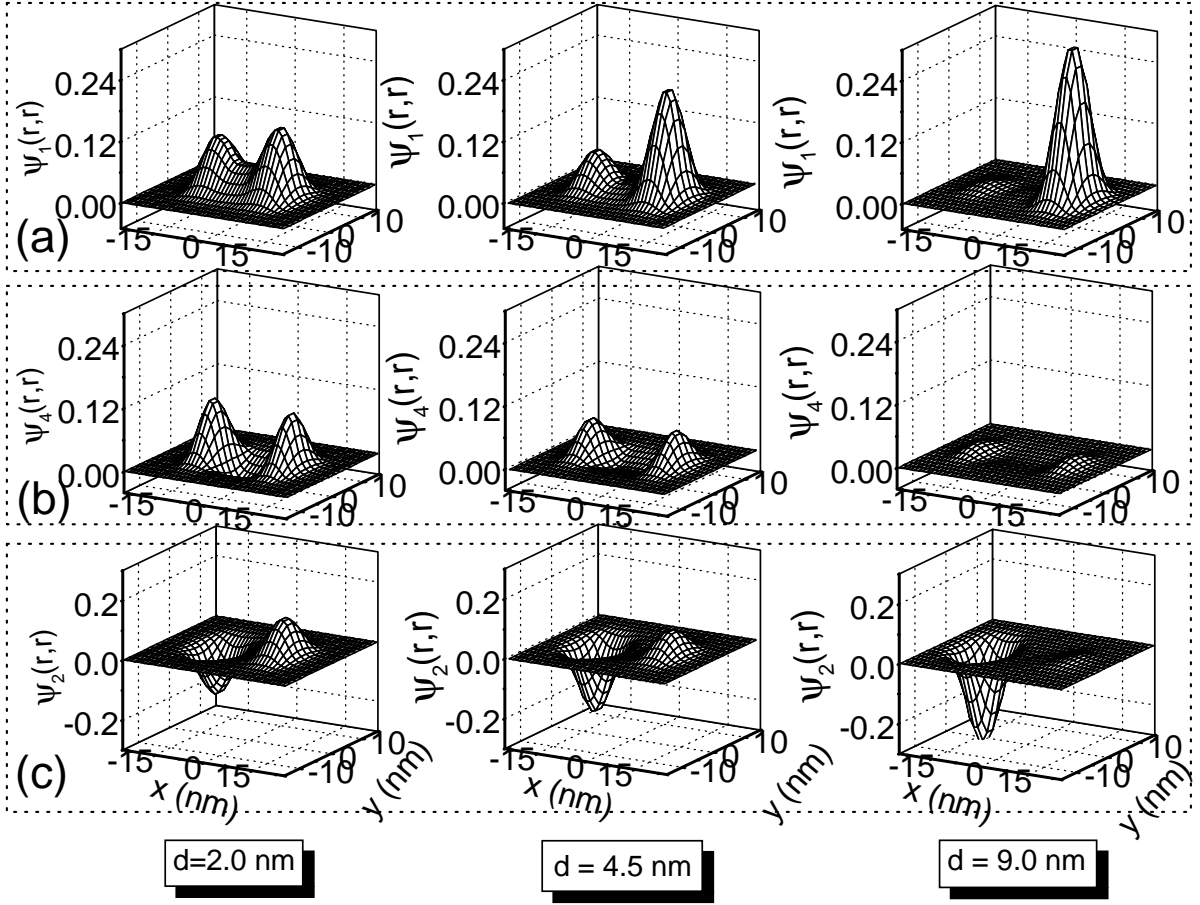}
\caption{(a) $\psi_1(r_e,r_h=r_e)$, (b) $\psi_4(r_e,r_h=r_e)$, and (c)
$\psi_2(r_e,r_h=r_e)$ of QDM in Fig. 6 for $d=2.0$, $4.5$, and $9.0$ nm,
respectively.} \label{FIG:polarization dis nonidentical}
\end{center}
\end{figure*}

Fine-structure splittings of three low-lying bright exciton levels are shown as
functions of $d$ for the two nonidentical dots in Fig.~\ref{FIG:ps
nonidentical}. Similar to the case of two identical dots, $\delta_1$ at larger
$d$ is negative, e.g., $\delta_1=-56$ $\mu$eV at $d=9.0$ nm. As $d$ decreases,
$\delta_1$ increases monotonically from negative values to positive values. At
$d\approx3.8$ nm, $\delta_1$ is zero. $\delta_4$ decreases from zero at larger
$d$ to $-150$ $\mu$eV at $d=1.5$ nm. The fine-structure splitting $\delta_2$ of
$\psi_2$ decreases from $-48$ $\mu$eV at $d=9.0$ nm to $-97$ $\mu$eV at $d=3.0$
nm. Similar to the definition in Eq.~\ref{inter intra}, the inter- and intradot
part of $\delta_2$ can also been given. Fig.~\ref{FIG:ps nonidentical}(b) shows
$\lambda_{1(2),\mathrm{inter}}$ and $\lambda_{1(2),\mathrm{intra}}$ as
functions of $d$. $\lambda_{1,\mathrm{intra}}$ is greatly changed from $-56$
$\mu$eV at $d=9.0$ nm to $54$ $\mu$eV at $d=1.0$ nm. In contrast to the
identical case, $\lambda_{1,\mathrm{inter}}$ is zero at $d=9.0$ nm and is
increased as $d$ is reduced. For $\psi_2$, $\lambda_{2,\mathrm{inter}}$
($\lambda_{2,\mathrm{intra}}$) is decreased from $-3$ $\mu$eV ($-45$ $\mu$eV)
at $d=9.0$ nm to $-37$ $\mu$eV ($-60$ $\mu$eV) at $d=3.0$ nm.

In nonidentical cases, $\psi_1(r_e,r_h=r_e)$ is almost localized in the larger
dot at $d=9.0$ nm as shown in Fig.~\ref{FIG:polarization dis nonidentical}(a)
and therefore $\lambda_{1,\mathrm{inter}}$ becomes zero at lager $d$ as shown
in Fig.~\ref{FIG:ps nonidentical}(b). Similar to the identical case,
$\psi_4(r_e,r_e)$ has two nodes along the $x$ axis and almost vanishes at
larger $d$. $\psi_2(r_e,r_e)$ is much localized in the smaller dot at $d=9.0$
nm as shown in Fig.~\ref{FIG:polarization dis nonidentical}(c) and therefore
$\lambda_{2,\mathrm{inter}}$, similar to $\lambda_{1,\mathrm{inter}}$, also
becomes zero at lager $d$. Since the symmetries of $\psi_1(r_e,r_e)$ and
$\psi_2(r_e,r_e)$ are opposite, $\lambda_{1,\mathrm{inter}}$ and
$\lambda_{2,\mathrm{inter}}$ have opposite signs, i.e.,
$\lambda_{1,\mathrm{inter}}>0$ and $\lambda_{2,\mathrm{inter}}<0$.

%$\psi(r_e,r_h=r_e)$ along $y=0$ axis of the exciton ground and second states
%are shown in Fig.~\ref{FIG:polarization dis nonidentical} for $d=6.0$ nm. In
%contrast to the identical case as shown in Fig. 5, $\psi(r,r)$ is much
%localized at 1-th and 0-th dot for the ground and second states, respectively.
%As a consequence, the exchange interaction in the nonidentical case is
%different to that in the identical one at the same inter-dot separation. For
%instance, at $d=9.0$ nm, $\delta_1$ is $-35$ and $-56$ $\mu$eV for the
%identical and nonidentical cases, respectively. exciton energy difference
%between two dots 1.76 meV.

\section{Summary}
We formulate a microscopic theory of exciton fine-structure splitting in
laterally-coupled QDM taking into account the Coulomb correlation between the
electron and hole. We choose the sizes of QDs resembling those grown in
experiments. In typical QDM, the electron and hole tunneling energies and the
direct Coulomb interaction are essential for the exciton levels and envelop
functions. As the inter-dot separation is reduced, fine-structure splitting of
the exciton ground state in QDM is largely increased while those of the second
and fourth states are decreased. At a proper separation, fine-structure
splitting of the ground state approaches zero. Fine-structure splittings of the
exciton ground and first-excited states could be divided into the intra- and
interdot exchange interactions, which are sensitive to the tunnel-coupling and
symmetry of QDMs. In symmetric QDMs, the inter-dot part is almost unchanged in
the range of few nanometer interdot separation, while in nonsymmetric QDMs the
inter-dot part is rapidly reduced to zero at lager inter-dot separation. In
both symmetric and nonsymmetric cases, the intra-dot part is greatly varied by
the tunnel-coupling, and the fine-structure splitting of the exciton ground
state could be varied by about hundred $\mu$eV. The fine-structure splitting of
the fourth state is greatly enhanced at smaller $d$ while approaches zero at
larger $d$. In summary, the study provides an efficient way of largely tuning
the exciton fine-structure splitting in semiconductor QDs, which is useful and
necessary for the fine-tuning by the external field.

%The variation of the fine-structure splitting as a function of inter-dot
%separation has different trends for exciton ground state and excited states.

\begin{acknowledgments}
Financial supports from NSF-China (Grant No. 10574077), the ``863" Programme
(No. 2006AA03Z0404) and MOST Programme (No. 2006CB0L0601) of China are
gratefully acknowledged.
\end{acknowledgments}

%\begin{references} \

\end{document}